\documentclass[12pt,a4paper]{article}
\usepackage{amsmath,amssymb,epsfig}
\usepackage{color,cite}

\renewcommand{\a}{\mathsf{a}}

\renewcommand{\P}{\mathbb{P}} 
\newcommand{\I}{\rm I}
\newcommand{\tr}{{\rm tr}}

\newcommand{\R}{{\mathbb R}}

\newcommand{\be}{\begin{equation}}
\newcommand{\ee}{\end{equation}} 
\newcommand{\bea}{\begin{eqnarray}}
\newcommand{\eea}{\end{eqnarray}}

\begin{document}

\title{On the Standard Model Group in F-theory}
\author{Kang-Sin Choi\footnote{email: kangsin@ewha.ac.kr}  \\ \normalsize \it
Scranton Honors Program, Ewha Womans University, Seoul 120-750, Korea}
\date{}
\maketitle

\begin{abstract}
We analyze the Standard Model gauge group $SU(3)\times SU(2) \times U(1)$ constructed in F-theory. The non-Abelian part $SU(3) \times SU(2)$ is described by singularities of Kodaira type. It is distinguished  to na\"ive product of $SU(3)$ and $SU(2)$, revealed by blow-up analysis, since the resolution procedures cannot be done separately to each group. The Abelian part $U(1)$ is constructed by obtaining a desirable global two-form harboring it, using `factorization method' similar to the decomposition method of the spectral cover; It makes use of an extra section in the elliptic fiber of the Calabi--Yau manifold, on which F-theory is compactified. Conventional gauge coupling unification of $SU(5)$ is achieved, without threshold correction from the flux along hypercharge direction. 
\end{abstract}

\newpage

\section{Introduction}

It has been well-known wisdom that the gauge group $SU(3) \times SU(2) \times U(1)$ of the Standard Model (SM) is far from arbitrary collection of simple and Abelian factors, because the matter fields transforming under this have very particular charge assignments. Grand Unified Theory (GUT) \cite{GUT} suggests that it is best understood by embedding the group to a series of the exceptional groups $E_n$, including $SU(5) \equiv E_4, SO(10) \equiv E_5,$ and $E_6$ \cite{En}. In this sense the SM group may be expressed as the unique, maximal $E_3 \times U(1)$ subgroup of $E_4$. These $E_n$ groups naturally occur in heterotic string and F-theory \cite{F,MV,BJPS,BHV,DW1,DW2,Tatar:2006dc,Hayashi:2008ba,Hayashi:2009ge,DW3,Choi:2009pv,Palti:2012aa}.

In this work, we analyze the singularities and the two-forms describing the Standard Model gauge group $SU(3) \times SU(2) \times U(1)$ in F-theory. The non-Abelian part can be described by conventional singularities of Kodaira type for the elliptic fiber in an internal manifold \cite{MV,Braun:2013cb,Beetal,Katz:2011qp}. Besides the sagacity that the GUT structure suggests, also in F-theory the desired light matter fields of the SM---not only the $\bf (3,2)$ representation but also $\bf (\overline 3,1)$ and $\bf (1,2)$---emerge on so-called matter curves \cite{KV,BHV,DW1,DW2}, {\em only if} the $SU(3) \times SU(2)$ is embedded in $E_6$ at least, because essentially the matter fields can only arise by branching the gauge multiplets in the {\em adjoint} representations of (local) unified groups. With the clue of various gauge symmetry enhancement directions, we can obtain the desired singularities for $SU(3)\times SU(2)$ by deforming the singularity of $SU(5)$, verified by matter curve structure, etc. \cite{CK,Choi:2010nf}. This will be reviewed and further analyzed by resolution process in Section 2.  Although the shape of the singularity was consistent, it has been obtained by applying various necessary conditions. The analyses given in Section 2 now provides a complete proof of it.
 
In constructing the gauge theory, the real problem has been an Abelian $U(1)$ group that is not obtained from a Kodaira-type singularity. We obtain the Abelian gauge field by expanding the three-form tensor of M/F-theory along a two-cycle \`a la Kaluza--Klein reduction. For the components of Cartan subalgebra of non-Abelian groups, we can automatically obtain such cycles by blowing-up the corresponding singularity. On the other hand it was quite difficult to find such a curve for $U(1)$ which is globally valid and has intersection numbers giving the desired charges of matter fields. Again, one hint can be embedding all the groups in a unified group, which would lead a two-form for Abelian group in the similar fashion for non-Abelian group. So-called $U(1)$-restricted model was the first successful method in describing such global $U(1)$ for $SU(5)\times U(1)$ by embedding the $U(1)$ group into $SU(2)$ \cite{u1restricted,Marsano:2011nn,Krause:2011xj}. However this heavily depends on a clever choice of ansatz and extending this to general gauge group, following the $E_n$ series was difficult. 
There have been an indirect derivation from spectral cover \cite{SpecCover} via heterotic--F-theory duality \cite{CH}, which gives a globally valid two-form in so-called stable degeneration limit, but it is useful in the case admitting the duality. Recently, Refs. \cite{MPW,Esole:2011sm}, introduced a `multiple section' method to introduce $U(1)$'s, essentially by finding a cousin of the element from Cartan subalgebra from a certain unified group (see also \cite{CH,Andreas:2004ja,Marsano:2011nn}). Such two-form naturally comes from more careful comparison between the elliptic equation and the spectral cover. In this paper, and especially in Section 3, we employ this method to obtain the $U(1)$ correctly describing hypercharge. As a byproduct, this method provides another proof to the expression of the singularity $SU(3) \times SU(2)$ in relation to the spectral cover. We employed this method because this is extended to any number of $U(1)$ sections and although the exceptional divisors may not be directly embedded into a larger group like $E_8$, the intersection relation can be embedded into and traced from such group. Another interesting direction is to find different sections making use of Mordell-Weil group generated by a single group element in the elliptic fiber and/or `tops' in toric geometry \cite{MWgroup,tops,CK}.

At the moment, there is no clue whether we have the Standard Model at the unification or string/F-theory scale without an intermediate GUT. Besides such {\it a priori} reason, the direct construction of the SM also has following practical merits. First, we do not need to turn on a flux in the hypercharge direction which gives rise to a threshold correction to the corresponding gauge coupling, ruining the coupling unification relation.  Second, some sector related to electroweak symmetry breaking is better understand if we have the unbroken group as the SM group, since some fields being footprints of GUT have nontrivial coupling to the Higgs sector \cite{Choi:2011te,e6gut}. Even we have the SM at the unification scale, still we have footprints of GUT such as the number of generation, since the string theory itself makes use of the unification relation \cite{CK,Choi:2010nf,Nilles:2009yd}. These two features have no analogy in heterotic string theory, since in F-theory we have two-step construction of gauge theory: constructing smaller group than $E_8$ and further symmetry breaking by $G$-flux. It controls the number of generation obeying a certain unification relation while the actual gauge symmetry is the smaller SM group.

\section{Non-Abelian factor $SU(3)\times SU(2)$}

We first construct a singularity for the non-Abelian algebra $SU(3)\times SU(2)$ of the Standard Model. As discussed in the introduction, it is not a mere product of simple algebras $SU(3)$ and $SU(2)$, but it should be along the chain of $E_n$ algebra\footnote{In this paper, we do not need to distinguish between the product of the group and the sum of the algebra, where the latter is more appropriate to our purpose, but we follow the traditional description.}. 

\subsection{Description by singularity} \label{sec:singularity}

We consider F-theory compactified on Calabi--Yau fourfold $Y$, which is an elliptic fiberation over a three-base $B$. The elliptic fiber is given as an hypersurface by an elliptic equation in the `Tate form'
\begin{equation} \label{ellipticeq}
 P \equiv -y^2 + x^3 + \a_1 xyz + \a_2 x^2 z^2 + \a_3 y + \a_4 x z^4 + \a_6 z^6 = 0
\end{equation}
in $\P^2_{[2,3,1]}$ fiber over $B$, having homogeneous coordinates $(x,y,z)$ with the respective weights indicated as subscripts, and all the parameters are also appropriate holomorphic sections over $B$ \cite{F,Katz:2011qp}. This can be regarded as a definition for our Calabi--Yau fourfold $Y$ as a hypersurface.

Tuning parameters $\a_i$ in some base coordinate, say $w$ defined with respect to a divisor of $B$
\begin{equation} \label{disclocus}
W: w=0,
\end{equation}
which should be already present from the construction of $B$, gives rise to singularities related to gauge symmetry on the worldvolume $W \times \R^4$. For instance, 
an $SU(5)$ singularity, split $\I_5$, is obtained from the table by Kodaira \cite{Beetal}.
\be \label{su5sg}
 \a_1 = b_5, \ \a_2 = b_4 w, \ \a_3 = b_2 w^2, \ \a_4 = b_2 w^3, \ \a_6= b_0 w^5, 
\ee
up to higher order terms in $w$. We have the discriminant of the elliptic equation (\ref{ellipticeq}) up to fifth order in $w$,
$$ \Delta = b_5^4 (b_0 b_5^2 - b_2 b_3 b_5 + b_3^2 b_4) w^5 + O(w^6),  $$
responsible for the gauge group $SU(5)$ \cite{F,MV,Beetal}.

Deforming the singularity by adding lower order terms in $w$ to $\a_i$'s, we have less severe singularity with smaller algebra. The claim in Refs. \cite{CK,Choi:2010nf} is that, the singularity for the $SU(3)\times SU(2)$ is given as,
\begin{align}
 \a_1 &= b_5 + O(w), \label{e3sg1} \\
 \a_2 &= b_4  w + O(w^2), \\
 \a_3 &= b_3 ( b_6 + w)w + O(w^3),\\
 \a_4 &= b_2 (b_6 + w)w^2 + O(w^4), \\
 \a_6 &= b_0  (b_6 + w)^2 w^3  + O(w^6). \label{e3sg6}
\end{align}
The discriminant takes the form
\begin{equation} \label{SMdiscr}
 \Delta =  b_5^3 P_{(3,2)}^2 P_{(\overline 3,1)} w^3  + P_{(3,2)} P_{30}
 w^4 + O (w^5)
\end{equation}
where the parameters are displayed in Table \ref{t:su32params}
and $P_{30}$ is a quite lengthy, non-factorizable polynomial in $b_i$, e.g. containing a term $2 b_3^2 b_4 b_5^4$.

We have a singularity so called Kodaira split $\I_3$ for $SU(3)$ located at $(x,y,w)=(0,0,0)$, which has orders ${\rm ord}(\a_1,\a_2,\a_3,\a_4,\a_6,\Delta)=(0,0,1,2,3,3)$ in $w$ \cite{Beetal,Katz:2011qp}. On the discriminant locus $W$ in (\ref{disclocus}), F-theory interprets it that we have the $SU(3)$ gauge theory \cite{F,MV}.
\begin{table} \begin{center}
\begin{tabular}{cccc}
\hline
name &  parameter & representation & symm. enhancement\\ \hline
 $P_{(3,2)}$  &  $b_6$ & ${\bf (3,2)}$ & $SU(5)$  \\
$P_{(\overline 3,1)}$  & $b_3^2 b_4 b_5 - b_2 b_3 b_5^2 + b_0 b_5^3 - b_3^3 b_6$ & ${\bf (\overline 3,1)} $ & $SU(4)\times SU(2)$\\
 $P_{(1,2)}$  &  $b_3^2 b_4 - b_2 b_3 b_5 + b_0 b_5^2 + b_2^2 b_6 - 4 b_0 b_4 b_6$& $\bf (1,2)$ & $SU(3) \times SU(3)$ 
 \\ \hline
\end{tabular}
\caption{Paramters of gauge symmetry enhancements.} \label{t:su32params}
\end{center} \end{table}
Setting $P_{(3,2)}=0$ enhances the discriminant to degree five, whereas $P_{(\overline 3,1)}=0$ does to degree four. The subscripts indicate the corresponding quantum numbers of unhiggsed fields in the branching, since $\bf (3,2)$ is regarded as off-diagonal component of the adjoint $\bf 24$ under the breaking $SU(5) \to SU(3)\times SU(2)$, while $\bf (\overline 3,1)$ is that of $\bf 15$ under  $SU(4)\to SU(3)$. The former symmetry enhancement shows that the actual group from the parameters (\ref{e3sg1})-(\ref{e3sg6}) is larger than $SU(3)$. It is because the parameters are specially {\em tuned up to} ${\rm ord}(\a_1,\a_2,\a_3,\a_4,\a_6)=(0,1,2,3,5)$, as the deformations of the $SU(5)$ singularity in (\ref{su5sg}).

To see the other part, we change the reference as
\begin{equation} \label{primrel}
 w' \equiv  w + b_6,
\end{equation}
defining a new divisor $W': w'=0$ of $B$.
The parameters become
\begin{equation}
 \begin{split} \label{smsurface2}
\a_1 &= b_5  + O(w'),\\
\a_2 &= b_4 (w' -b_6) + O(w^{\prime 2}),\\
\a_3 &= b_3 (w'- b_6)w' + O(w^{\prime 3}), \\
\a_4 &= b_2 (w'- b_6 )^2 w' + O(w^{ \prime 4}), \\
\a_6 &= b_0 (w'- b_6 )^3 w^{\prime  2} + O(w^{ \prime 6}).
 \end{split}
\end{equation}
The discriminant has the form 
\begin{equation} \label{discsu2part}
 \Delta =  \left( b_5^2 - 4 b_4 b_6  \right)^2 P_{(3,2)}^3 P_{(1,2)} w^{\prime 2} +  P_{(3,2)}^2  P'_{30}
 w^{\prime 3} + P_{(3,2)} P_{36} w^{\prime 4} + O(w^{\prime 5})
\end{equation}
where the parameters are shown in Table \ref{t:su32params}
and $P'_{30},P_{36}$ are non-factorizable polynomials containing respectively $3 b_3^2 b_4 b_5^4, -3 b_3^2 b_4 b_5^5 b_6$. From the observation that ${\rm
  ord}(\a_1,\a_2,\a_3,\a_4,\a_6,\Delta)=(0,0,1,1,2,2)$ in $w'$ we see at $(x,y,w')=(0,0,0)$ there is the Kodaira singularity, split $\I_2$ for $SU(2)$. Again we have the $SU(2)$ gauge theory localized on the locus $W'$. The parameter $w'$ is distinguished to $w$ by the relation (\ref{primrel}) via the parameter $b_6$, which is the section depending on the base coordinate.
We see the parameter $P_{(3,2)} = b_6$ again since $(\bf 3,2)$ is also charged under this and vanishing of which enhance the gauge symmetry to $SU(5)$, in which limit we do not distinguish between $w$ and $w'$. Also $P_{(1,2)}=0$ enhances the symmetry $SU(2) \to SU(3)$. It is clear that our singularity describes the maximal semisimple algebra $SU(3)\times SU(2)$ embedded in $SU(5)$.

\subsection{Resolution} \label{sec:resol}

We resolve the $SU(3) \times SU(2)$ singularities following the Tate algorithm \cite{Ta,Beetal,Krause:2011xj}. This resolution shall reveal nontrivial algebraic structure.
Neglecting higher order terms in $w$, the elliptic equation is
\begin{equation} \label{su3su2}
 P= -y^2 + x^3 + b_5 xyz + b_4 w x^2 z^2 + b_3(b_6 + w)w y z^3 + b_2(b_6 + w) w^2 x z^4 + b_0 (b_6 + w)^2 w^3  z^6.
\end{equation}

First we resolve the $\I_3$ part located at  $(x,y,w)=(0,0,0)$. We introduce another affine coordinate $e_1$ of a $\P^1$ curve such that
$$ (x,y,w) = (x_1 e_1, y_1 e_1, w_1 e_1),$$
and forbid the simultaneous vanishing $x_1 = y_1 = w_1 = 0$. Then the original singularity is only accessed by $e_1=0$.
The lowest order terms in $e_1$ have common factor $y_1$, so still the point $(e_1,y_1)=(0,0)$ is again singular. To have smooth resolution of $Y$, we blow-up again
\begin{equation} \label{e2scaling}
 (e_1,y_1) = (e_1' e_2, y_2 e_2),
\end{equation}
or equivalently $(x,y,w) = (x_2 e_1' e_2, y_2 e_1' e_2^2, w_2 e_1' e_2)$, and remove $e_1' = y_2=0$.
Then the lowest order terms now in $e_2$ have no common factor and the resolution procedure terminates. From now on we drop the subscripts in $x,y,w$ and the prime in $e_1'$, and so on, if there is no confusion. The resulting polynomial is
\begin{equation} \begin{split}
 \tilde P =&  e_1^2 e_2^3 \big[ x^3  e_1- y^2  e_2  + b_5 xyz + b_4  e_1 w x^2 z^2  + b_3 ( b_6 +  e_1 e_2 w) w y z^3  \\
&+ b_2 (b_6 +  e_1 e_2 w )  e_1 w^2  x z^4 + b_0 (b_6  + e_1 e_2 w )^2  e_1 w^3  z^6 \big].
\end{split}
\end{equation}

Next, we go to the $SU(2)$ part, by going to the primed coordinates using the relation (\ref{primrel}), however, now posessing the form
\begin{equation}
 b_6 +  e_1 e_2 w \equiv   w',  \label{chcoord}
\end{equation}
(Here $w$ means $w_2$ in the above (\ref{e2scaling})). We obtain
\begin{equation} \begin{split}
 \tilde P' =&   x^3  e_1^3 e_2^3 - y^2  e_1^2 e_2^4  + b_5 e_1^2 e_2^3 xyz +  b_4  e_1^2 e_2^2 (w'  - b_6) x^2 z^2  + b_3 w' (w' - b_6) e_1  e_2^2 y z^3  \\
 &+ b_2 w'  (w'-b_6)^2 e_1 e_2  x z^4 + b_0 w^{\prime 2} (w'-b_6)^3   z^6 .
\end{split}
\end{equation}
As before this describes $\I_2$ singularity at $(x,y,w')=(0,0,0)$. We want blow up there by introducing another coordinate $e$ such that  $w' \to w'e$.
$$ (x,y,w') \to (xe, ye, w'e). $$
\begin{equation} \begin{split} \label{Pprime}
\hat P' =& e^2 \big[x^3  e_1^3 e_2^3 e - y^2  e_1^2 e_2^4  + b_5 e_1^2 e_2^3 xyz +  b_4  e_1^2 e_2^2 (w'e  - b_6) x^2 z^2  + b_3 w' (w' e- b_6) e_1  e_2^2 y z^3  \\
 &+ b_2 w'  (w' e-b_6)^2 e_1 e_2  x z^4 + b_0 w^{\prime 2} (w'e-b_6)^3   z^6 \big].
\end{split}
\end{equation}
This is the standard resolution of $\I_2$ singularity, found in e.g. Ref \cite{MPW}. It seems in this primed coordinates, we would have more singularities such as $(e_2,w') = (0,0)$. Shortly we see, it turns out we have no more. We come back to the original coordinates, now by the modified relation
\begin{equation} \label{coordrel}
 w'  e- b_6  =  e_1 e_2 e_0. 
\end{equation}
Then the equation becomes
\begin{equation} \begin{split} \label{hatP}
 \hat P =&  e_1^2 e_2^3 \big[ x^3 e^3 e_1- y^2 e^2 e_2  + b_5 e^2 xyz + b_4 e_0  e_1 e^2 x^2 z^2 +
 b_3 (e_0 e_1 e_2 + b_6) e_0 e y z^3  \\
&+ b_2 (e_0 e_1 e_2 + b_6) e_0^{2}  e_1 e x z^4 + b_0 (e_0 e_1 e_2 + b_6)^2 e_0^{3}  e_1    z^6 ]
\end{split}
\end{equation}
We have changed the name $w$ to $e_0$, since the divisor $e_0 =0$ plays the role of the extended root of $SU(3)$ below.

The two resolution procedures should commute or should not prefer the order. Indeed, 
we see it is, since we can write the overall result as
$$ (x, y, w, w') \to (xe e_1 e_2, ye  e_1 e_2^2, e_0   e_1 e_2, w' e) $$
where the coordinate $w$ is changed to $e_0$ for later convenience. Nevertheless two resolutions affect each other due to the constraint (\ref{coordrel}). 
The resolution procedures can be equivalently re-expresed in terms of the scaling in Table \ref{t:scaling}. Besides the definition $Z$ for the $\P^2_{2,3,1}$, we have introduced three new coordinates $e_1,e_2,e$ and three scaling relations $E_1,E_2,E$. 

\begin{table} \begin{center}
\begin{tabular}{cccccccccc}
\hline
 & $x$ & $y$ & $z$ & $e_1$ & $e_2$ & $e$ & $e_0$ & $w'$ \\ \hline
$Z$ & 2 & 3 & 1 & 0 & 0 & 0 & 0  &0  \\
$E_1$ & $1$ & $1$ & 0 & $-1$ & $0$ & $0$ & $1$ & 0  \\ 
$E_2$ & $1$ & $2$ & 0 & $0$ & $-1$ & $0$ & $1$ & 0 \\
$E$ & $ 1$ & $1$ & 0 & 0 & 0 & $-1$ & $0$ & 1\\ \hline
\end{tabular}
\caption{Scaling relations from the definition of the exceptional divisors $e_1,e_2$ and $e$. We do not remove $w'$ by scaling, although it scales covariantly, but just constrained by (\ref{coordrel}).} \label{t:scaling}
\end{center} \end{table}

Once the scalings are established, for instance $E_1$ in Table \ref{t:scaling} means  $(x,y,z,e_1,e_2,e,e_0) \to (\lambda x, \lambda y,  z, \lambda^{-1} e_1, e_2,e, \lambda e_0)$, we should exclude some points $x=y=e_0=0$ as explained above. A combination of two scalings always gives a new scaling in a different guise and we see it has the structure of ideal. So we introduce the Stanrey--Reisner (SR) ideal, containing such data, generated by
\begin{equation} \begin{split} \label{SRideal}
\{ xyz, & xy e_0, ye_1, x e_0 e_2, y z e,x ze_2  , z e_1 e_2, xy w'\}  \cup \{ (ze_1,ze)  \text{ xor } x e_0  \}\cup \{ ze_2  \text{ xor } y e_0 \} \\
\{ e_1 w'& \text{ xor } e e_0 \} \cup \{ e_2 w' \text{ xor } y e e_0\}\cup \{ z e e_2  \text{ xor } e_0 w' \}  \cup \{ x e_2 w' \text{ xor } ( e e_0, e_1 e) \}\cup \{ x e_1 w ' \text{ xor } e_2 e \}  
  \end{split}
\end{equation}
where in each curly parenthesis, we can choose one of the elements (xor means exclusive or), corresponding to a particular triangulation of the toric diagram \cite{Blaszczyk:2011hs}. We {\em must} choose $z e_1$ and $z e_2$ to have four-dimensional Lorentz vector components for the Cartan subalgebras that will be related to $e_1$ and $e_2$, which we see below. What we choose here are $e e_0, e_2 w', e_0 w', e_1 e,$ and $x e_1 w'$, some of which generated by others. Finally we have 
\begin{equation} \label{SRchoice}
  \{ xyz , xye_0,  ye_1, x e_0 e_2, ze_1, ze_2, ze, xyw', e_0e, e_1 e,e_0 w', e_2w'\}.
\end{equation}

\subsection{Intersections} \label{sec:intersec}

We hereafter consider divisors of the Calabi--Yau manifold $\hat Y$ defined by $\hat P=0$. 
Vanishing loci of the blow-up coordinates $e_i$ define exceptional divisors $E_i$. Explicitly we have
\begin{align}
 E_1 
  &:  e_1= 0 = - e_2 + b_5 x + b_3 b_6 e_0  , \quad \{ ye_1, ze_1, e e_1 \} \label{e_1} \\
 E_2  
 &: e_2 = 0= x^3 e^3 e_1 + b_5 e^2 xy + b_4 e_0 e_1 e^2 x^2   \\
 & \qquad + b_3 b_6 e e_0 y + b_2 b_ 6 e e_0^2 e_1  x  + b_0 e_0^3 e_1 b_6^2  ,  \quad \{ z e_2  \}  \nonumber \\
 E_0 
 &: e_0 = 0= x^3 e_1 - y^2 e_2 + b_5 xyz=0, \quad \{ e_0 e \}  \\
 E &: e=0=-y^2 e_2^4 + b_5 e_2^3 xy- b_4 b_6 e_2^2 x^2 \nonumber \\
  & \qquad - b_3 b_6 w' e_2^2 y + b_2 b_6^2 e_2 w' x - b_0 b_6^3 w^{\prime 2}, \quad \{e e_1,ze \} \label{Edivisor} \\
 W'  &:  w'=0=  x^3 e_1 e - y^2 + b_5 xyz - b_4 b_6 x^2 z^2 . \quad \{ e_0 w', e_2 w' \} \label{wprime}
\end{align}
In the every second line, we simplified the relation using the SR ideal (\ref{SRchoice}). 
The divisors $E_1, E_2,E_0$ are the objects in the $SU(3)$ part, so we obtain them from $\hat P_T$ after performing proper transformation $e_1^2 e_2^3 \hat P_T = \hat P$  in (\ref{hatP}) and we obtain $E,W'$ of the $SU(2)$ from $e^2 \hat P'_T = \hat P'$ in (\ref{Pprime}). In particular the divisor $E$ has dependence on the coordinate $w'$ that cannot be eliminated by the constraint (\ref{coordrel}), so it is not able to be compared with the $E_1, E_2, E_0$ in the $SU(3)$ part. However, due to the constraint (\ref{coordrel}), $E$ does not have common intersection with them. For the same reason, 
we simply decouple the factor $e_1^2 e_2^2$ in the equation $\hat P'_T|_{w'=0}=0$ to define $W'$, since the constraint (\ref{coordrel}) forbids vanishing of the either factor. Using the constraint $e_2 = -b_6$, we may massage the equation (\ref{Edivisor}) to a fancier form, which in fact becomes important later.

A complete intersection of $E_i$ with two arbitrary divisors $D_a$ and $D_b$ in $\hat Y$ gives a $\P^1$ curve. Since $D_a$ and $D_b$ are arbitrary, an intersection number of two such $\P^1$ curves is given by the number of common solutions to the equations $E_i = E_j = \hat P= 0$:
\begin{align}
 E_1 \cdot E_2 = 1 &: \quad e_1 = e_2=  b_5 x + b_3 b_6 e_0  =0, \\
 E_1 \cdot E_0 = 1 &: \quad e_1 = e_0 =-e_2 + b_5 x = 0, \\
 E_2 \cdot E_0 = 1 &: 
 \quad e_2 = e_0 = e_1 + b_5 y = 0. \quad \{ x e_0 e_2, e_0 e, e_2 e \},
\end{align}
where the dot product notation is understood.
Each equation has one solution in $x,y,$ and/or $e_i$, assuming that $b_i$'s are all nonzero.
This completes the $SU(3)$ root relations via McKay correspondence that the intersection numbers corresponding to the minus of the Cartan matrix of the algebra. The above intersections are also expressed as \cite{Grimm:2011tb}
\begin{equation} \label{mckay}
 \int_{\hat Y} E_i \wedge E_j \wedge D_a \wedge D_b = - A_{ij} \int_B W \wedge D_a \wedge D_b, \quad E_i \in \text{Cartan subalgebra}, 
\end{equation}
where $A_{ij}$ is the Cartan matrix and $D_a,D_b$ are divisors in $B$ (whose pullback to $\hat Y$ is omitted without confusion). 

These exceptional divisors $E_1, E_2, E_0$, thus the corresponding roots in the $SU(3)$ algebra, are disconnected to the rest of divisors $E$ and $W'$ of $SU(2)$, since the constraint does not allow simultaneous vanishing of $e_i$ and $e$, or of $e_i$ and $w'$ for each $i=1,2,0$. 

For $E$ and $W'$, we have two solutions in $y/x$ to
\be \label{selfint}
 E \cdot W' = 2 :\quad e=w'= y^2  - b_5 x y + b_4 b_6 x^2 = 0, \quad \{ e z , e_0 e, e_1 e, e_2 w'\}
 \ee
consistent with the affine (roughly, extended) Dynkin diagram of $SU(2)$. If there is only one solution, the discriminant of (\ref{selfint}) becomes 
\begin{equation} \label{discrofye2}
 b_5^2 - 4 b_4 b_6 = 0,
\end{equation}
which destroy the $O(w^{\prime 2})$ term in the discriminant (\ref{discsu2part}) of the elliptic equation. We will assume otherwise in what follows.

\subsection{Matter curves and symmetry enhancement}

In Section \ref{sec:singularity}, we have studied various gauge symmetry enhancements by analyzing the discriminant. 
In Table \ref{t:su32params}, each equation $ P_f = 0$ defines a codimension one curve of the $SU(3)$ surface $e_0=0$ and/or the $SU(2)$ surface $w'=0$. Since we can interpret it as that there are light matter fields $f$ localized on the $P_f=0$, we call it as the matter curve. Here we further analyze the matter curves from properties of the exceptional divisors resulting from the resolution.

\paragraph{Matter curves for $\bf (3,2)$}

We first analyze the matter curve $P_{(3,2)} =0$ for $\bf (3,2)$. On this locus, there is local gauge symmetry enhancement to $SU(5)$. Here we will see this is reflected by further degeneration and rearrangement of the exceptional divisors.

The equations for exceptional divisors become
\begin{align}
 E_1 \to E_{1A} &:   \label{E1A} \quad
  - e_2 + b_5 x  = 0 = e_1, \\
 E_2 \to E_{2x} \cup E_{2E} \cup E_{2B} &: \quad x e^2 (x^2 e e_1 + b_5 y + b_4 e_0 e_1  x)  = 0 = e_2,   \\
 E_0 \to E_{0C} &: \quad x^3 e_1 - y^2 e_2 +b_5 xy z = 0 = e_0, \\
 E \to E_{E2}  &:\quad  e_2^3 y (-y e_2 + b_5 x) = 0 = e, \label{EE2} 
\end{align}
where we have degeneration of $E_2$ and we renamed the divisors accordingly. We may find $E_{2E}:e_2=e=0$ from $E$ as well, and now the previously disconnected part can communicate via this. It satisfies the modified constraint (\ref{coordrel}) now read as $e = e_2$ using the SR elements $e_0e$ and $e_1e$. This locks only possible divisor from $E$ to $E_{2E}$, among  seemingly possible elements e.g. $E_y: e=y=0$.

Consequently, the only nontrivial intersections are
\begin{align}
 E_{0C} \cdot E_{1A} = 1 &: \quad e_0 = e_1 = - e_2 + b_5 x = 0, \\
 E_{1A} \cdot E_{2x} = 1 &: \quad e_2 = e_1 = x = 0, \\
 E_{2E} \cdot E_{2x} = 1 &: \quad e_2 = e = x = 0,  \label{E2E2x}\\
 E_{2E} \cdot E_{2B} =1 &: \quad e_2 = e = b_5 y + b_4  x = 0, \quad \{ e_0 e\}  \label{E2E2A} \\
 E_{2B} \cdot E_{0C} = 1 &: \quad e_2 = e_0 =  e_1 + b_5 y = 0. 
\end{align}
The rest intersection numbers are zero. For instance, $E_{2x} \cdot E_{0C} =0: e_0 = e_2 = x = 0$ forbidden by the SR element $x e_0 e_2$.
In particular there is no intersection between $E_{2x}$ and $E_{2B}$ since the required equations $e_2 = x=y=0$ are forbidden by the SR element $xy$ which is inherited by $xy w'$ with $w'=0$.

Altogether these relations give rise to the extended Dynkin diagram the locally enhanced gauge symmetry $SU(5)$. However globally the unbroken gauge symmetry on $\hat Y$ still remains $SU(3)\times SU(2)$.
Although the divisor $E$ had no intersections to $E_2$ before the symmetry enhancement, now their degenerated daughters do have nonzero intersections. This can be tracked to general factors $e$ appearing in the divisors, resulted from the resolution of the $\I_2$ part that cannot be {\em separately} done with respect to the $\I_3$ part.

\begin{table}
\begin{center}
\begin{tabular}{cc} \hline
divisor & weight \\ \hline
$E_{1A}$ & $[-2,1;0]$ \\
$E_{2x}$ & $[1,-1;1]$ \\
$E_{2E}$ & $[0,0;-2]$ \\
$E_{2B}$ & $[0,-1;1]$ \\
$E_{0C}$ & $[1,1;0]$ \\ \hline
\end{tabular}
\end{center} \label{t:weights}
\caption{$SU(3) \times SU(2)$ weights of the exceptional divisors in Dynkin basis. }
\end{table}

Also we can calculate the $SU(3)\times SU(2)$ weight of the divisors (\ref{E1A})-(\ref{EE2})  in Dynkin basis, as $[E_i \cdot E_1, E_i \cdot E_2; E_i \cdot E],$ shown in Table \ref{t:weights}.
At the intersection or matter curve $b_6=0$, we have local gauge symmetry enhancement which explains the emergence of a light field with quantum number $\bf (3,2)$. Their six components and weights are displayed in Table \ref{t:threetwo}. As expected two roots of $SU(3)$ are played by $E_{1A}$ and $E_{2x}+E_{2E}+E_{2B}$, and the root of $SU(2)$ is $E_{2E}$. One can easily see that this is the only representation whose components are only effective divisors.

\begin{table}
\begin{center}
\begin{tabular}{cc} \hline
divisor & weight \\ \hline
$E_{2B}+E_{0C}$ & $[1,0;1]$ \\
$E_{2B}+E_{0C} + E_{2E}$ & $[1,0;-1]$ \\
$E_{2B}+E_{0C}+E_{1A} $ & $[-1,1;1]$ \\
$E_{2B}+E_{0C}+E_{1A} +E_{2E} $ & $[-1,1;-1]$ \\
$E_{2B}$ & $[0,-1;1]$ \\
$E_{2B}+E_{2E} $ & $[0,-1;-1]$ \\
\hline
\end{tabular}
\end{center}
\caption{The components of the matter representations $\bf (3,2)$, at the matter curve $b_6 = 0$.} \label{t:threetwo}
\end{table}

\paragraph{Matter curves for $\bf (\overline 3,1)$}
There are other gauge symmetry enhancement directions, according to Table \ref{t:su32params}.
We have $SU(3) \to SU(4)$ symmetry enhancement, yielding light matter $\bf (\overline 3,1)$ at the matter curve
$ P_{(\overline 3,1)} = 0. $
For example solving in $b_6$ and restoring appropriately $b_6$ again, we have a splitting of $E_2$.
\begin{equation} \begin{split}
    E_2 \to E_{2D} \cup E_{2F} &: e_2 = 0 = b_3^{-2} b_5^{-1} b_0 (b_3 b_6 e_0 + b_5 e x) \\
 & \times \left[ b_3 b_5 b_6 e_0^2 e_1 + b_5 (b_2 b_3 - b_0 b_5 ) e e_0 e_1 x +b_3^2 e( b_5 y+   e e_1 x^2) \right] .
\end{split} \end{equation}
We have $E_0 \cdot E_1=1$ as before and in addition we have intersections of new divisors
\begin{align}
 E_{1} \cdot E_{2D} = 1  &: \quad e_1 = e_2 = b_5 x + b_3 b_6 e_0 = 0, \\
 E_{1} \cdot E_{2F} = 0 &: \quad e_1 = e_2 = b_3^4 b_5 = b_5 x + b_3 b_6 = 0, \quad \{ e_1 e, e_1 y\} \\
 E_0 \cdot E_{2D} =0 &: \quad e_0 =e_2 = x =0, \quad \{  x e_0 e_2 \} \\
 E_0 \cdot E_{2F} = 1 &: \quad e_0 = e_2 = e_1 + b_5 y =0. \quad \{ x e_0 e_2 \}
\end{align}
The representations are calculated in Table \ref{t:threeone}, where we must take the highest weight $[-1,1;0]$, not $[0,1;0]$.

\begin{table}[t] \begin{center}
\begin{tabular}{cc} \hline
divisor & weight \\ \hline
$ E_0 + E_1 + E_{2F} $ & $[-1,1;0]$ \\
$ E_{2F} $ & $[0,-1;0]$ \\
$E_1 + E_{2F}$ & $[-2,0;0]$ 
\\ \hline
\end{tabular} 
\qquad
\begin{tabular}{cc} \hline
divisor & weight \\ \hline
$E_G $ & $[0,0;1]$ \\
$E_G+ E$ & $[0,0;-1]$ 
\\ \hline
\end{tabular}
\end{center}
\caption{The components of the matter representations $\bf (\overline 3,1)$ at the matter curve $P_{(\overline 3,1)}= 0$, and $\bf (1,2)$ at the matter curve $P_{(1,2)}= 0$.} \label{t:threeone}
\end{table}

\paragraph{Matter curves for $\bf (1,2)$}

Another symmetry enhancement direction is $P_{(1,2)}= 0$, shown in Table \ref{t:mequation}. We use the equation for the divisor $E$ after applying the constraint (\ref{coordrel}), now reads as $e_2 = -b_6$,
\begin{equation} \label{Edivagain}
 E: e= 0 = b_6 y^2 + b_5 xy+ b_4 x^2 + b_3 w' y + b_2 w' x + b_0 w^{\prime 2},
\end{equation}
after dropping $-b_6^3$.
It may degenerate into two divisors
\begin{equation} \label{EGEHdivs}
 E_G \cup E_H: e= 0 = (p w' + q x + r y)(s w' + u x + v y)  ,
\end{equation}
where 
$$ b_0= ps,\ b_2= pu+qs,\ b_3=sr+pv,\ b_4=qu,\ b_5=ru+qv,\ b_6=rv,$$
satisfying the relation $P_{(1,2)}=0$ in a highly nontrivial way. We can always solve six parameters $p,q,r,s,u,v$ for as many $b_i$'s. In fact, setting $w'=1$ makes (\ref{Edivagain}) to be identical to spectral cover equation, in which the condition for local factorization (\ref{EGEHdivs}) is precisely the requirement for the existence of the $\bf (1,2,20)$ representation of $SU(3)\times SU(2)\times SU(6) \subset E_8$ \cite{Hayashi:2008ba}. This is another evidence that the divisor equation for $E$ (\ref{Edivagain}) should be obtained from $\hat P'_T$ not from $\hat P_T$ .

Accordingly, each $E_G$ or $E_H$ provides a representation for $\bf (1,2)$, shown in Table \ref{t:threeone}. We may check the intersection relations in the same way
\be
 E_G \cdot E_H = 1: e = p w' + q x + r y = s w' + u x + v y = 0
\ee
where we have nontrivial solution to the last two equations in $x$ and $y$ if 
\be
 (ur-qv)^2 = b_5^2 -4 b_4 b_6 \ne 0,
\ee
which we have already met in (\ref{discrofye2}) and assumed nonvanishing.
As is
\begin{equation}
 E_G \cdot E = E_G \cdot (E_G + E_H) = -2 + 1 = -1,
\end{equation}
so holds the same relation for $E_H$, by the symmetry that we have no qualitative difference between $E_G$ and $E_H$.
We see also there is no distinction between $E_G$ and $E_H$ group theoretically, since $\bf (1,2)$ is a self-conjugate representation. But this may be subject to Freed--Witten global anomaly. This factorization and intersection structure of $E$ indirectly suggests that we should obtain $E$ from the primed coordinates $\hat P'$ not $\hat P$.

\section{Abelian factor $U(1)$}

The work by Mayrhofer, Palti and Weigand \cite{MPW}, following that of Esole and Yau \cite{Esole:2011sm}, introduced a systematic way to obtain arbitrary number of $U(1)$ gauge fields with desired gauge quantum numbers, which we follow here (see also Ref. \cite{CH}, which has similar structure making use of heterotic/F-theory duality). The idea is to introduce a new section than zero section in the elliptic fiber, by tuning the coefficients of the elliptic equation. This is very similar and indeed related to the factorization of the spectral cover to which we can relate the group elements. Then we have a conifold singularity at the new section. Resolving this new singularity will give rise to a new section $S$ having wished intersection structures with the existing resolution divisors for the desirable gauge quantum numbers.

\subsection{More global sections from factorization}

We need one-forms or gauge fields $A_1$ of Abelian symmetries for either Cartan subalgebra of non-Abelian groups or just $U(1)$ groups. To realize them, we require harmonic two-forms $w_2 \in H^{1,1}(\hat Y)$ by which the M/F-theory three-form tensor is expanded as
\begin{equation} \begin{split} \label{C3expansion}
 C_3 &= \sum A_1 \wedge w_2 \\
   & = A_1^{e_1} \wedge w_2^{e_1} + A_1^{e_2} \wedge w_2^{e_2} + A_1^{e} \wedge w_2^e + A_1^Y \wedge w_2^Y + \cdots.
   \end{split} \end{equation}
Here, $w_2^{e_i}$ are the dual two-forms to the divisors $E_1,E_2$ or $E$ of $\hat Y$ obtained from the blowing-ups in the previous section. 

We need two kinds of requirements for $w_2$'s. (i) Each $A_1$ should have desired gauge quantum number, so the Poincar\'e-dual divisor of the paired $w_2$ in $\hat Y$ should have appropriate intersections with other divisors. (ii) Every $A_1$ should be seven-dimensional vector field, which restricts the index structure of the components of the pared $w_2$ \cite{DW1}.
For a resolved Kodaira singularity, the blown-up cycles $w_2^{e_i}$ automatically possess Requirement (i), seen from the intersection structure, seen in Section \ref{sec:intersec}. However there is no Kodaira singularity for an Abelian symmetry, or would-be-related $\I_1$ singularity is actually smooth. So we cannot do the resolution as in Section \ref{sec:resol}. We will see shortly that the desired two-form is obtained from another kind of singularity under a more special condition. The goal of this section is to find $w_2^Y$ harboring the hypercharge satisfying Requirement (ii). 

An important hint is the relation between the elliptic equation and spectral cover equation. The latter in a sense shows the algebraic relation in more suggestive form, since the coefficients are directly related to combinations of weight vectors. The relation is best seen by restricting the former on the hypersurface
\be \label{Xdef}
  X \equiv e_0 e_1^2 e x^3 - ( e_0 e_1 e_2 + b_6) y^2 = e(e_0 e_1^2  x^3 - w' y^2) = 0.
\ee
It will be convenient to define $t$ as
\be \label{tdefinition}
x e_1 t \equiv y, \quad \{e_1 y,xy \}
\ee
serving as a well-behaved holomorphic coordinate. It is because the condition $t=0$ means $y=0$, and the opposite also holds because the SR elements indicated in (\ref{tdefinition}) forbid $x=0$ and $e_1=0$. So from now on we use $t = y e_1^{-1} x^{-1}$, under which (\ref{Xdef}) becomes
\begin{equation}
   e_0 e x- ( e_0 e_1 e_2 + b_6) t^2.
\end{equation}
Putting this to $\hat P$, we obtain
\begin{equation} \label{notfactorized}
 {\hat P}\big|_{X=0} = e_0^{-3} e_1 e^2 w^{\prime 2} (b_0 e_0^6 z^6 + b_2 t^2 e_0^4z^4 + b_3 t^3 e_0^3z^3 + b_4 t^4 e_0^2z^2 + b_5 t^5 e_0 z + b_6 t^6).
\end{equation}
Unless $b_6=0$, $E_0$ has no intersection to $X$ so we do not care about the factor $e_0^{-3}$, otherwise there is a cancellation and no overall factor in $e_0$ remains. Besides the prefactor, the polynomial in $z$, in the parenthesis, is nothing but the spectral cover equation, whose relation is discussed in Ref. \cite{MPW,Hayashi:2008ba}. The F-theory compactification requires a global section in the elliptic fiber, which is at $(x,y,z)=(1,1,0)$ in our case and usually called zero section $Z$.  In the spectral cover description, the vanishing sum of five distinguished points, for the unimodular group $SU$, should be translated to the absence of $z^5$ term, with our choice of the zero section.  


In the spectral cover description, we have obtained a globally valid $U(1)$ by `decomposition' \cite{CH}. Its first procedure is tuning of parameters (\ref{e3sg1})-(\ref{e3sg6}) as follows \cite{MPW,CK,DW3,DW1,Hayashi:2008ba,Marsano:2009gv}. 
\be \label{tuning15}
\begin{split}
 b_0 &= a_0 d_1, \quad b_2 = a_0 d_2 + a_1 d_1,\quad b_3 = a_0 d_3 + a_1 d_2, \\
 b_4 &= a_0 d_4 + a_1 d_3, \quad b_5 = a_0 d_5 + a_1 d_4, \quad b_6 = a_1 d_5.
\end{split} \ee
In other words,
\begin{align}
 \a_1 &= (a_0 d_5  + a_1d_4), \label{su5sg1} \\
 \a_2 &= (a_0 d_4 + a_1 d_3 ) w, \\
 \a_3 &= (a_0 d_3 + a_1 d_2 )(a_1 d_5 + w)w, \\
 \a_4 &= (a_0 d_2 + a_1 d_1  )(a_1 d_5 + w)w^2, \\
 \a_6 &= a_0 d_0 (a_1 d_5 + w)^2 w^3. \label{su5sg6}
\end{align}
And the absence of $\a_5$ or $b_1$ should be translated as a constraint 
$$0 = a_0 d_1 + a_1 d_0.$$
Then the Tate polynomial becomes factorized form 
\begin{equation} \begin{split} \label{factorization14}
  \hat P \big|_{X=0} &=e_0^{-3} e_1 e^2 w^{\prime 2} (a_0 e_0 z + a_1 t)(d_0 e_0^5 z^5 + d_1 e_0^4 z^4 t+ d_2 e_0^3 z^3 t^2 + d_3 e_0^2 z^2 t^3 + d_4 e_0 z t^4 + d_5  t^5).\\
  &\equiv Y_1 Y_2.
 \end{split}
\end{equation}
This will become the $S[U(1)\times U(5)]$ spectral cover equation in the heterotic side.
However to ensure this $U(1)$ be global, we require {\em another} global section other than the zero section \cite{MPW,CK}.
In our case, indeed we have the new section is at $X = Y_1 = 0$. Since this section is related to the `$U(1)$ part' $Y_1$ or the parameter $a_1/a_0$, we may expect this section has appropriate structure for the hypercharge. 
\begin{table}[t]
\renewcommand{\arraystretch}{1.3}
\begin{center} \begin{tabular}{|c|c|c|} \hline
matter &  parameters  \\ \hline
$P_X$ & $a_1$   \\  \hline
$P_{q_\circ}$ & $d_5$  \\ \hline
$P_{u^c_\circ}$ & $a_1^3 d_2 + a_0 a_1^2 d_3 + a_0^2 a_1 d_4 + a_0^3 d_5 
$
 \\
\hline
$P_{d^c_\circ}$ &
 \parbox{8cm}{ $a_0 a_1 d_2 d_3 d_4 + a_0^2 d_3^2 d_4 + a_1^2 d_0 d_4^2 - a_0 a_1 d_2^2 d_5$ \\ $
 - a_0^2 d_2 d_3 d_5 + 2 a_0 a_1 d_0 d_4 d_5 + a_0^2 d_0 d_5$
 } 
\\ \hline
$P_{l_\circ}$ &
\parbox{10cm}{
 $ a_1^5
d_5d_0^2-a_1^3d_5a_0^2d_2d_0-2a_1d_5a_0^4d_0d_4-3a_1^2d_5a_0^3d_0d_3$ \\
$+a_0^3d_0a_1^2d_4^2 +a_0^5d_0d_5^2+a_0^4
d_2a_1d_4d_3-a_0^5d_2d_5d_3+a_1^3d_0d_4a_0^2d_3$\\
$+a_0a_1^4d_0 d_4d_2 +a_0^5d_4d_3^2+a_0^4a_1
d_3^3+2a_0^3a_1^2d_3^2d_2+a_0^2a_1^3d_3d_2^2$
} 
 \\ \hline
$P_{e^c_\circ}$ &  $-2 d_1 a_1^4 +d_2 a_0 a_1^3- d_3 a_0^2 a_1^2 +d_4 a_0^3
 a_1^2 - d_5 a_0^4 $ 
\\ \hline
  \end{tabular}
\caption{Defining equation for the matter curves.}
\label{t:mequation}
\renewcommand{\arraystretch}{1}
\end{center}
\end{table}

The factorization structure (\ref{factorization14}) may be expressed in another way \cite{MPW}
\be \label{binomial1}
 \hat P_T = X Q -  Y_1 Y_2 = 0,
\ee
by introducing a polynomial $Q$, holomorphic in $z$ and $t$. This leads to a conifold singularity at 
\be \label{conifold}
 X = Q= Y_1 = Y_2 = 0,
 \ee
which is of higher codimension than one.
We blow up $Y_1=Q=0$ by introducing a $\P^1$ with homogeneous coordinates $(\lambda_1, \lambda_2)$ such that \cite{Esole:2011sm}
\begin{equation} \label{res}
 Y_1 \lambda_2 = Q \lambda_1, \quad Y_2 \lambda_1 = X \lambda_2. 
 \end{equation}
The original singularity (\ref{conifold}) gives unconstrained $\lambda_1$ and $\lambda_2$, which means it is replaced by the $\P^1$. Away from the singularity, we recover the original equation (\ref{binomial1}) by solving $\lambda_1$ and $\lambda_2$. In effect, the equations in (\ref{res}) have redefined the Calabi--Yau space, which we denote as $\hat Y$ again by abusing the notation, as a hypersurface in the new ambient space including the $\P^1$. 
Then, we obtain an an extra section than zero section as the following divisor in $\hat Y$
\be \label{Sdivisor}
 S: \lambda_1=0, 
 \ee
which forces $\lambda_2$ be nonzero and gives $Y_1 = X = 0$ from (\ref{res}). The lesson from the spectral cover \cite{CH} tells us that $Y_1=0$ is related to the hypercharge $U(1)$ as a subset of the commutant  $S[U(1)\times U(5)]$ to the SM group in $E_8$.
The desired candidate for the hypercharge $(1,1)$-form is therefore Poincar\'e-Hodge dual to the threefold $S$ in $\hat Y$, up to some correction we should consider below.

Next, we consider Condition (ii) below Eq. (\ref{C3expansion}), for $A_1$ being a Lorentz vector.
A natural method is given in Ref. \cite{MPW}, which we follow here with similar notations. Such forms $w$ should satisfy following constraints
\begin{align}
 \int_{\hat Y} w &\wedge D_a \wedge D_b \wedge D_c = 0,  \label{dconstraint} \\
 \int_{\hat Y} w &\wedge Z \wedge D_a \wedge D_b = 0,  \label{zconstraint}
\end{align}
from which the two indices of $w$ have one leg on the elliptic fiber and the other leg on the base $B$. Here the divisors $D_a,D_b,D_c$ of $\hat Y$ are pullbacks of arbitrary divisors in $B$. It is far from trivial for the Cartan subalgebra elements $E_1$ and $E_2$ to satisfy the relation (\ref{zconstraint}), without choosing the SR ideal (\ref{SRideal}).  For $w$ we find the desired linear combination
\begin{equation} \label{Shioda}
 w_Y =  S - Z - \bar {\cal K} +  a_1 + \sum t_i E_i 
\end{equation}
where $\bar {\cal K}$ is the canonical class of the base $B$, $a_1$ is the divisor defined by $a_1=0$, and the coefficients $t_i$ will be determined later. The general such process is called Shioda map \cite{Shioda}.
The condition (\ref{dconstraint}) is satisfied using the relation
$$ \int_{\hat Y} S \wedge D_a \wedge D_b \wedge D_c  = \int_{\hat Y} Z \wedge D_a \wedge D_b \wedge D_c = \int_B D_a \wedge D_b \wedge D_c   $$ 
since both $S$ and $Z$ are global section. The next condition (\ref{zconstraint}) is satisfied by (\ref{Shioda}) thanks to the following. First, the intersection of $Z: z =0$ means $Y_1 = a_1 t$ among the defining equation of $S$. Using that $xyz$ is an element SR ideal, the only nontriviality for intersection $S \cdot Z$ comes from $$ \int_{\hat Y} S\wedge Z \wedge D_a \wedge D_b = \int_B a_1 \wedge D_a \wedge D_b. $$
Also  the adjunction formula states
$$ \int_{\hat Y} Z\wedge Z \wedge D_a \wedge D_b = -\int_B 
 \overline {\cal K}  \wedge D_a \wedge D_b. $$

\subsection{More symmetry enhancement}

The procedure described in the previous subsection introduces two new things: one is the new section $S$ and the other is further factorization of the matter curve. 
We calculate the intersection of the divisors with $S$ under various symmetry enhancement conditions on matter curves. From this, we see that, although $S$ did not originate from the Cartan subalgebra of the $SU(5)$ singularity or Kodaira $\I_5$, it plays exactly the same role for the fourth generator of it, other than existing generators $E_1,E_2,E$.

\paragraph{Matter curves for $\bf (3,2)$}

Under the parametrization (\ref{su5sg1})-(\ref{su5sg6}) the matter curve equation $P_{(3,2)} = 0$ in (\ref{SMdiscr}) further factors as
$$ P_{(3,2)} = P_q P_X = 0, $$
with different intersection structures of exceptional divisors for each factor shown in Table \ref{t:mequation}. We can directly compute the intersections as, omitting $\lambda_1 = D_a=0$, 
\begin{align}
  S \cdot E_{1A,q}= 0&:\quad  e_1 = y = -e_2 +a_1 d_4 x = d_5= 0,\quad  \{ y e_1 \}\\
 S \cdot E_{2E,X} =1 &: \quad e_2 = e = x = a_1 =   0, \quad \{x e_0 e_2, e_0 e, e_1 e\} \label{ScdotE2E}\\
 S \cdot E_{2E,q} =1 &: \quad e_2 = e = Y_1  =d_5 =  0, \\
S \cdot E_{2B,X} =1&: \quad e_2 = e_0 =  e_1 + a_0 d_5y  = a_1=  0, \quad \{ x e_0 e_2, e_0 e\} \\ 
S \cdot E_{2B,q} =0&: \quad e_2 = X = Y_1 = x^2 e e_1 + a_1 d_4 y + (a_1 d_3 + a_0 d_4) e_0 e_1 x  = d_5   = 0, \label{Sdot2Bq} \\
S \cdot E_{2x,q} =0 &: \quad e_2 = x = y = d_5 =  0. \quad \{xy\}
\end{align}
In calculating the intersection  of $Y_1 = a_0 e_0 + a_1 t$ with $x=0$ or $e_1=0$, the definition of $t$ is not valid so we need to restore its original form $a_0 e_0 e_1 x+ a_1 y$. 

We should remember that although we have local gauge symmetry enhancement, still on $\hat Y$ the gauge symmetry is $SU(3)\times SU(2)\times U(1)$, whose basis corresponds to $E_1,E_2,E$. Thus we have a definite product between those with $S$. We should have a definite intersection $S \cdot E_1=S \cdot E_{1A,q} =0$ so we should also have
$$S \cdot E_{1A,X} = 0. $$
As before, we have an invariant
\begin{equation} \label{SdotE2}
 S \cdot E_2 = S \cdot (E_{2x} + E_{2E} + E_{2B}) =1
\end{equation} 
calculated from the matter curve $q$. Since $S \cdot E_2$ should be independent of the decomposition, therefore we should have the same value for the $X$ and we have 
$$S \cdot E_{2x,X}=-1.$$ 
Also indirectly we can obtain the value. From the definition of the extended root $E_0$, we have linear dependence relation $E_{0C} +E_{1A}+ E_{2x} + E_{2E} + E_{2B}=0$, fixing,  for both $X$ and $q$,
 $$S \cdot E_{0C} = -1.$$

\paragraph{Matter curves for $\bf (\overline 3,1)$}

Now we go to the case of $\bf (\overline 3,1)$. In this case our factorization is 
$$ P_{(\overline 3,1)}= P_{u^c} P_{d^c} = 0.
$$
where each factor is again displayed in Table \ref{t:mequation}.
Also omitting $\lambda_1= D_a=0$, we have
\begin{align}
 S \cdot E_{2D,u^c} = 1: & \quad e_2 = b_3 b_6 e_0 + b_5 e x = X = 0, \label{S2Duc} \\
 S \cdot E_{2D,d^c} = 0: & \quad e_2 = b_3 b_6 e_0 + b_5 e x = X = Y_1 = 0,  \label{S2Ddc} \\
 S \cdot E_{2F,u^c} =  0: & \quad e_2 = b_3 b_5 b_6 e_0^2 e_1 + b_5 (b_2 b_3 - b_0 b_5 ) e e_0 e_1 x +b_3^2e( b_5 y+   e e_1 x^2)  = X = Y_1=0 \\
 S \cdot E_{2F,d^c} =  1: & \quad e_2 = b_3 b_5 b_6 e_0^2 e_1 + b_5 (b_2 b_3 - b_0 b_5 ) e e_0 e_1 x +b_3^2 e( b_5 y+   e e_1 x^2) = X = 0.
\end{align}
While the constraint $P_{u^c}=a_1^2 b_3 + a_0^2 b_5 = 0$ makes the conditions in (\ref{S2Duc}) automatically solve the equation $Y_1 = 0$, it is not in the case of $d^c$ curve in (\ref{S2Ddc}). The same situation holds for $E_{2F}$. The rest of intersection is the same as in the previous case $S \cdot E_0 = -1, S \cdot E_1 =0$.

There are still no matter curve for $\bf (1,2)$ for the factorization (\ref{tuning15}); For generic $a_i$'s and $d_i$'s, we cannot solve the six parameters in (\ref{EGEHdivs}).

\paragraph{Matter curve for $\bf (1,2)$}

With the factorization at $X=0$, one of $E_G$ and $E_H$ has a solution in the form
\begin{align}
 E_G:& \quad p+qx+ry = (a_0 e_0 z+ a_1 t)(g_0 e_0^2 z^2+ g_1 e_0 zt+ g_2t^2 )=0,
\end{align}
with the constraint $a_0 g_1+a_1 g_0 = 0$. This is regarded as the definition of $E_G$ from now on, and the other part $E_H$ is untouched. Thus the modified defining equation of $E_G$ contains the factor $Y_1=0$ and the condition is redundant. This is the reason why we have no further factorization of $P_{(2,1)}$.
Therefore we have the intersection structure
\begin{align}
 S \cdot E_{G} = 1&:  \quad P_{(2,1)} = X = Y_1 = \lambda_1 = D_a = 0,\\
 S \cdot E_{H} = 0&.
\end{align}
At the same time can distinguish the lepton doublet from the down-type Higgs doublet by extra $U(1)$ quantum numbers than hypercharge.

\paragraph{Matter curve for $\bf (1,1)$}

After the factorization to obtain the hypercharge symmetry $U(1)_Y$, we have a new charged singlet $e^c: {\bf (1,1)}_1$ under the SM group $SU(3)\times SU(2)_L\times U(1)_Y$, shown in Table \ref{t:mequation}. On the matter curve $P_{e^c}=0$, we have gauge symmetry enhancement $U(1)_Y \to SU(2)_R$ so that the resulting $SU(3)\times SU(2)_L \times SU(2)_R$ is still a subgroup of $SO(10)$ \cite{Choi:2011ua}. 

We note that $P_{e^c}=0$ is contained in the complete intersection between $Y_1$ and $Y_2$. Around here, the fiber equation (\ref{binomial1}) locally has a binomial structure of a deformed Kodaira $\I_2$ equation $xy=z_1 z_2$, which describes nothing but this $SU(2)_R$ gauge symmetry \cite{u1restricted,MPW,Esole:2011sm}. Therefore the small resolution gives the $\P^1$ fiber (\ref{conifold}) over the locus $P_{e^c}=0$, which we now call $S'$. This $S'$ will be related to the weight vector for $e^c$. Away from the intersection $P_{e^c}=0$ in the base $B$, its fiber described by (\ref{conifold}) is already a $\P^1$, which we call $E'$. These two have McKay correspondence of the affine $SU(2)_R,$ namely  $S' \cdot E' = 2. $ And we can show that $S$ intersects the entire fiber $S'+ E'$ at a single point \cite{MPW}. Therefore $S'$ provides the desirable intersection giving the correct hypercharge of $e^c$,
\begin{equation}
 S \cdot S'  = -1.
\end{equation}

\subsection{Hypercharge generator from the embedding}

In the previous section, we have studied the intersections of the new divisor $S$ in (\ref{Sdivisor}) with various divisors, or, to be more precise, the intersection numbers between their $\P^1$ fibers in the sense of (\ref{mckay}). Although on various loci $P_f=0$ each of the divisors $E_1,E_2,E$ may further degenerate into many, we can recollect the results in terms of the intersections among $E_1,E_2,E$ and $S$. For example, the relation (\ref{SdotE2}) may be recollected as $S \cdot E_2 =1 $ since this relation is independent on any specific locus $D_a = D_b = 0$ on which we calculate.

Therefore we  summarize the result as follows.  We have McKay correspondence of intersections of the $\P^1$ fibers
\be \label{su5extintersections}
 \bordermatrix{~ & E_1 & E_2 & S & E \cr
                  E_1 & -2 & 1 & 0 & 0 \cr
                  E_2 & 1 & -2 &  1 & 0 \cr
                  S & 0& 1& -2 & 1 \cr
                  E & 0 & 0 & 1 & -2                              
                  } = - A_{SU(5)}
\ee
being the minus of the Cartan matrix of $SU(5)$. The divisor $S$ provides the `fourth root' of $SU(5)$. This is a good news, since the hypercharges are correctly given to the fields when we choose $S$ as the generator with a suitable normalization. The divisors $E_1,E_2,E$ may be blown down to zero size to recover nonabelian singularity $SU(3)\times SU(2)$. However the divisor $S$ cannot be blown down maintaining the factorization (\ref{factorization14}), since we cannot allow the conifold singularity with higher codimension. Therefore, at best we can have the gauge group $SU(3)\times SU(2) \times U(1)$, not the full $SU(5)$.

However the Poincar\'e--Hodge dual two-form to the divisor $S$ is not exactly what we want as hypercharge generator, since we need a disconnected $\omega_Y \simeq S$ from the other group $SU(3)\times SU(2)$ as
\be
 \int_{\hat Y} w_Y \wedge E_i  \wedge D_a \wedge D_b = 0 \quad \text{ for } E_1, E_2, E.
 \label{cartandir}
\ee
We may form a linear combination of $S$ with $E_i$'s to have the desired property. This is to find the coefficients $t_i$ in (\ref{Shioda}) in the Shioda map \cite{Shioda} done in the following mnemonics. Take the inverse Cartan matrix $A^{-1}$ of the enhanced gauge symmetry of rank $r+1$. The Dynkin basis is defined to be and provides a convenient orthogonal relation between roots $\alpha_i$ and weights $w^i$ in a group under consideration
$$ \alpha_i \cdot w^j = \delta_i^j, \quad a_i = \sum_j A_{ij} w^j, $$
where the Cartan matrix provides the product metric and the sum is done over all the weights in that algebra.
The symmetry breaking is described by deleting the $j$th node of the Dynkin diagram and the resulting unbroken symmetry with Cartan matrix being the one with without the $j$th row and $j$th column. From the orthonormality relation, what we need here is to take $j$th row of the {\em inverted} Cartan matrix as the coefficients $t_i$ of linear combinations of root divisors $E_i$.  In our case 
$$
 A_{SU(5)}^{-1}  = \frac15 \begin{pmatrix} 4 & 3 & 2 & 1 \\
 3 & 6 & 4 & 2 \\
 2 & 4 & 6 & 3 \\
 1 & 2 & 3 & 4 \\ \end{pmatrix}.
$$
The symmetry breaking $SU(5) \to SU(3)\times SU(2) \times U(1)$ is done by removing 3rd row (and removing the extended root of the $SU(5)$). Therefore, we take the third row $(2,4,6,3)$, neglecting the overall normalization $1/5$, to obtain $2,4,6,3$ for coefficients of $E_1,E_2,S$, and $E$, respectively. This always guarantee the integral charges under this $U(1)$.
We finally have the hypercharge generator
\begin{equation} \label{hypercharge}
 w_Y =  -\left[S - Z - \bar {\cal K} - a_1 +  \frac{1}{6} (2E_1 + 4 E_2 + 3E)\right],
 \end{equation}
with the overall normalization chosen according to the conventional charge. Applying this to any component $E_f$ of each field $f$ gives the hypercharge $Y_f = \int_{E_f} w_Y$ as
\begin{align}
\ Y_q =  \frac16,\ Y_X = -\frac56 ,\ Y_{u^c} =-\frac{2}{3}, \ Y_{d^c} =\frac13,\ Y_{l}=- \frac12,\ Y_{e^c} = 1.
\end{align} 
It is highly nontrivial that every component has different inner product with $6S$ and $2E_1+4E_2+3E$, but their sum is always the same, as it should be for the components in a same multiplet.

\subsection{Further factorization}

The model we have been building so far cannot be realistic for some reasons below. 
\begin{enumerate}
\item Even with a `1+5' factorization giving the hypercharge $U(1)$, we could not obtain the massless field with the quantum number $\bf (2,1)$ for generic parameters $a_i$'s and $d_i$'s, since we have no solution to the equation for the desired weights (\ref{EGEHdivs}). We may hope that further tuning of these parameters may solve the problem. 
\item
The spectrum so far, listed in Table \ref{t:mequation}, cannot take into account the Higgs fields. By further factorization, we hope we can distinguish up and down Higgses by their localization on different matter curves. 
\item To have four dimensional chiral spectrum, we have to turn on $G$-flux. If we turn on the universal $G$-flux along the entire `$SU(5)$' part\footnote{For convenience we call this $SU(5)$ part as a commutant group of the $SU(3) \times SU(2) \times U(1)$ in $E_8$, just borrowing the nomenclatures of spectral cover construction. The other commutant in this case is the hypercharge $U(1)_Y$ since the abelian group commutes to itself.} we have partial unification relation of $SU(5)$. That is, the SM field belonging to the same representation of $SU(5)$ has the same number of generations among themselves. For example $n_q= n_{u^c}=n_{e^c}$ from $\bf 10$ where $n_f$ is the number of generations of a matter field $f$. If we want more strong unification relation, we may turn on $G$-flux along smaller part than $SU(5)$, for instance $SU(4)$, which gives a larger commutant for the unification relation, for instance of $SO(10)$. 

Note that always the unbroken group here is the SM group $SU(3)\times SU(2)\times U(1)_Y$, purely determined by the tuning of the parameters (\ref{tuning15}) of the elliptic equation, regardless the choice of $G$-flux.  
\end{enumerate}
In spectral cover construction, it was  shown that factorization with the spectral cover $S[U(3)\times U(1)\times U(1)\times U(1)]$ is most realistic \cite{Choi:2011te}, and the same applies to our F-theory version. 

As before, we seek extra sections as subset of the variety $X=0$. So we will require the factorization of the elliptic equation in the form (we will drop the factor $e_0$ and $z$ for simplicity)\footnote{In what follows we have new definitions on $Y_i$'s and $Q$, etc. and they are not related to similar ones in the previous section.}
$$
 \hat P|_{X=0} = (a_0+ a_1 t)(b_0 + b_1 t)(d_0 + d_1 t)(f_0 + f_1 t + f_2 t^2 + f_3 t^3) \equiv Y_1 Y_2 Y_3 Y_4 
$$
with the constraint
$$ a_1 b_0 d_0 f_0 + a_0 b_1 d_0 f_0 + a_0 b_0 d_1 f_0 + a_0 b_0 d_0 f_1 = 0. $$
Rewriting this again as $XQ =Y_1 Y_2 Y_3 Y_4$ by introducing a holomorphic polynomial $Q$, the above four factors of $Y_i$'s give rise to singularities. We know we shall have three $U(1)$'s so we introduce as many $\P^1$'s with homogeneous coordinates $(\lambda_1,\lambda_2),(\mu_1,\mu_2),(\nu_1,\nu_2)$ and choose a resolution as
\begin{equation} 
   Y_1  \lambda_2    = Q \lambda_1 , \quad 
    Y_2  \mu_2 =  \mu_1 \nu_2, \quad
 Y_3 \lambda_1 \mu_1 \nu_1 =  \lambda_2 \mu_2, \quad
  Y_4  \nu_2=  X \nu_1.
\end{equation}
The resulting manifold is smooth since the Jacobian has the maximal rank. We are content to verify that at least locally this reduces to the binomial resolution for three factors, shown in Ref. \cite{Esole:2011sm}. When we have locally $Y_2 \simeq 1$, we have $\mu_2 = \mu_1 \nu_2$. Plugging them, the relations agree as
$$ Y_1 \lambda_2 = Q \lambda_1, \quad Y_3 \lambda_1 \nu_1 = \lambda_2 \nu_2, \quad Y_4 \nu_2 = X \nu_1. $$ 
When $Y_3 \simeq 1$, we have $\mu_2 = \lambda_1 \mu_1 \nu_1 / \lambda_2$ on one patch $\lambda_2 \ne 0$ and we reproduce
$$ Y_1 \lambda_2 = Q \lambda_1, \quad Y_2 \lambda_1 \nu_1 = \lambda_2 \nu_2, \quad Y_4 \nu_2 = X \nu_1. $$
This also holds good on the other patch $\lambda_1 \ne 0$.

Consequently, we have new exceptional hypersurfaces containing the sections $X=Y_i=0$ for $i=1,2,3$
\begin{align}
 S&:    \lambda_1 = \mu_2  =\nu_2= 0 \Longrightarrow X=Y_1 = 0, \\
 S_X &:  \lambda_2 = \mu_1 = \nu_1 = 0 \Longrightarrow X=Y_2 = Q = 0, \\
 S_Z &: \lambda_ 2 = \mu_2 = \nu_2 =0 \Longrightarrow X = Y_3 = Q = 0.
\end{align}
The two extra $U(1)$ charges we call $X$ and $Z$. This $S$ divisor has essentially the same definition as that in the previous factorization (\ref{Sdivisor}), having the same group and Lorentz properties.
As before, the newly found divisors $S_X$ and $S_Z$ provide `missing' Cartan subalgebra of $SU(5)$ or $SO(10)$, respectively. We can recycles the $U(1)$ generators $w_Y$, since the latter satisfies all the requirement of Cartan subalgebra (\ref{cartandir}) and the whole generators satisfy desirable conditions (\ref{dconstraint}), (\ref{zconstraint}). Thus we find the new generators with normalization
\begin{align}
 w_X =& 5(S_X - Z - \overline {\cal K} - b_1) + 2 E_1 + 4 E_2 + 6 E + 3 (-6w_Y),\\
 w_Z =& 4(S_Z - Z - \overline {\cal K} - d_1) + 2 E_1 +4  E_2 +6 E + 5
   w_X +3 (-6w_Y),
\end{align}
where the factor $-6$ in front of $w_Y$ is due to the special fractional convention of hypercharge. Since $S_X$ and $S_Z$ are going to belong to $SO(10)$ and $E_6$, respectively, the coefficients are also found from the inverted Cartan matrices
$$
 A_{SO(10)}^{-1} = \frac14 \begin{pmatrix}  4 & 4 & 4 & 2 & 2 \\
 4 & 8 & 8 & 4 & 4 \\
 4 & 8 & 12 & 6 & 6 \\
 2 & 4 & 6 & 5 & 3 \\
 2 & 4 & 6 & 3 & 5 \\
 \end{pmatrix},
 \quad
 A_{E_6}^{-1} = \frac13 \begin{pmatrix}
  4 & 5 & 6 & 4 & 2 & 3 \\
 5 & 10 & 12 & 8 & 4 & 6 \\
 6 & 12 & 18 & 12 & 6 & 9 \\
 4 & 8 & 12 & 10 & 5 & 6 \\
 2 & 4 & 6 & 5 & 4 & 3 \\
 3 & 6 & 9 & 6 & 3 & 6 \\
 \end{pmatrix}
$$
Further generalization is straightforward.
The spectrum of the fields and the corresponding charges are shown in Table. 1 in Ref. \cite{Choi:2011te}.

\section{Comment on gauge coupling unification} \label{sec:gaugecoupling}

With localization of each gauge theory on a complex surface $S_4$ in $B$, a part of eight dimensional worldvolume, we have the following field theory limit having dimensional reduction \cite{BHV,DW1}
\begin{equation} \label{lowE}
 - \frac{e^{-\phi}}{(2 \pi \alpha')^4} \int_{S_4 \times \R^4} d^{8} xF^2_{8D} = -\frac{\rm Vol \,\it S_{\rm 4}}{4 g^2_{\rm YM}} \int_{\R^4} d^4x F^2_{4D} + \cdots,
 \end{equation}
with the vacuum expectation value of the dilaton $e^\phi$ becoming string coupling.
In the IIB string theory limit, this ${\rm Vol}\,S_4$ is interpreted as the effective volume of the cycle wrapped by dynamical severbranes with both NSNS and RR charges. 

The volumes of $S_{(3)}$ and $S_{(2)}$ respectively spanned by the $SU(3)$ locus $W \equiv E_0: e_0=0$ (in $B$) and the $SU(2)$ locus $W': w'=0$ are related, using (\ref{su5extintersections}).
\begin{equation} \label{twovolumes}
 \begin{split}
{\rm Vol} S_{(3)}&=  \frac12 \int_{S_{(3)}} J \wedge J = \frac12 \int_B W \wedge J \wedge J= -\frac12 \int_{\hat Y} E_2 \wedge S \wedge J \wedge J \\
&= - \frac12 \int_{\hat Y} E \wedge S \wedge J \wedge J = \frac12 \int_B W' \wedge J \wedge J = \frac12  \int_{S_{(2)}} J \wedge J = {\rm Vol} S_{(2)}. 
\end{split} \end{equation}
where $J$ is the K\"ahler form of $\hat Y$. Therefore we have the same worldvolume for these two non-Abelian gauge groups. Here the calibrated geometry plays a role: the effective volumes are given by intersection numbers, not depending on scaling factors of the coordinates. 

The gauge coupling of the hypercharge $U(1)$ can be readily determined in the relation to the enhanced group such as $SU(5)$. It should be a global limit $b_6 = 0$, i.e. not local gauge symmetry enhancement on matter curves
\begin{equation} \begin{split}
 \hat P|_{b_6=0} =&  e_1^2 e_2^3 \big[ x^3 e^3 e_1- y^2 e^2 e_2  + b_5 e^2 xyz + b_4 e_0  e_1 e^2 x^2 z^2 +
 b_3 e_0^2 e_1 e_2  e y z^3  \\
 &+ b_2 e_0^3 e_1^2 e_2 e x z^4 + b_0 e_0^5 e_1^3 e_2^2     z^6 ]
 \end{split}
\end{equation}
with the tuned parameters $b_i$ in (\ref{tuning15}).
It is happy to see that in this limit, the two-cycles $e_0, e_1, e_2$ are identical to those in the standard resolution of $SU(5)$ singularity $\I_5$ (See, e.g. \cite{MPW}, after renaming $e_2 \to e_4$). 
The woldvolume is provided by the divisor $W_0 : e_0 = \hat P_T|_{b_6=0}=0$, which is the same as $W$ of the $SU(3)$. Thus
\begin{equation}
{\rm Vol} S_{(5)} =  \frac12 \int_{S_{(5)}} J \wedge J = \frac12 \int_B W_0 \wedge J \wedge J  =\frac12 \int_B W \wedge J \wedge J = {\rm Vol} S_{(3)}.
\end{equation}
The fact that $W, W', W_0$ have the same volume is obvious since the $SU(3)$ and the $SU(2)$ are obtained by deforming $SU(5)$ singularity. It is not affected by another deformation arising from the resolution $S$ of the conifold singularity. The volume of the $\P^1$ fiber of $S$ cannot be nonzero so there cannot be unbroken $SU(5)$. Nevertheless the gauge couplings are unaffected by the volume of this $\P^1$ and in the low energy limit, we just have heavy $X,Y$ gauge multiplets.

In this limit, $S$ provides the Cartan subalgebra element related to hypercharge, as seen in the relation (\ref{su5extintersections}) thus
$$ - \frac{1}{4g^2} \tr F_{SU(5)}^2 = - \frac{1}{4g^2} ( \tr F_{SU(3)}^2 + \tr F_{SU(2)}^2 + \tr F_{U(1)}^2),  $$
where we defined the gauge field as matrix valued $A_M = A_M^a t^a, \tr t^a t^b = \frac12 \delta^{ab}$. In particular, from the Cartan matrix (\ref{su5extintersections}), the generator $S$ is related to the Cartan element with $t = \frac{1}{\sqrt{60}}{\rm diag}(2,2,2,-3,-3)$. 
The two-cycle $w_Y$ is just a modification of that of $S$, and the linear transformation within the same group $SU(5)$ (otherwise even the definition (\ref{hypercharge}) does not make sense) is just a transformation not affecting the gauge coupling.
Thus the gauge coupling of the hypercharge $U(1)_Y$ should be related by group theory of the unified group $SU(5)$ rembedding $SU(3) \times SU(2) \times U(1)_Y$, in the standard way. Normalizing the $U(1)$ charge of the $e^{\rm c}$ to be $1$, we fix the coupling as
$$ g^2 = g_3^2 = g_2^2 = \frac35 g_Y^2, $$
with the weak mixing angle at this string theory scale
$$ \sin \theta_W^0 = \frac{g_Y^2}{g_2^2 + g_Y^2} = \frac38,  $$
consistent with the observation. For any $U(1)$ having an embedding to a certain GUT, we may use this method, however it is an open question whether every $U(1)$ obtainable in F-theory has such embedding.

To this coupling relation, we have threshold correction, if there is a nontrivial $G$-flux along a certain $U(1)$ direction. When we construct the SM group at the string scale, we should not turn on $G$-flux along the hypercharge direction if we want it to be gauge symmetry. For GUT such as $SU(5)$, we may break it by turning on $G$-flux without breaking gauged hypercharge \cite{BHV,DW2}. Other $U(1)$ symmetries constructed for the realistic model building, it is desirable to broken down by $G$-flux. Then by St\"uckelberg mechanism, the corresponding gauge boson acquires mass and the symmetry becomes global.
There are also threshold correction to it from the flux \cite{DW2,Blumenhagen:2008aw,Heckman:2010pv}.

\section{Conclusion}
We analyzed the Standard Model gauge group $SU(3)\times SU(2) \times U(1)$, and also its accompanying matter fields, constructed in F-theory, using resolution procedures. The non-Abelian part $SU(3) \times SU(2)$ is described by the singularities of Kodaira type, which locally looks like $\I_3$ and $\I_2$, as nontrivial deformation of the $SU(5)$ singularity $\I_5$. They are respectively supported at different divisors $w=0$ and $w'=0$, which are related by a coordinate transformation (\ref{coordrel}), nevertheless described by single elliptic equation (\ref{hatP}). The resolution analysis revealed that the SM group should be distinguished to na\"ive product of $SU(3)$ and $SU(2)$, since the two groups are connected by coordinate transformations and the blowing-ups cannot be done independently. On the matter curves, there are gauge the symmetry enhancements to various unified groups, and the exceptional divisors from different simple groups mix in some particular way, yielding matter fields having desired charges. This desirable feature is present only if the SM group is embedded in $E_n$ series group. 

The Abelian part $U(1)$ is obtained  by `factorization method' making use of an extra section in the elliptic fiber of an internal manifold. At a particular restriction $X=0$, the factorization of the elliptic equation is related to gauge symmetry enhancement in certain group direction. The resolution at the conifold singularity originating from this factorization gives rise to the two-form harboring the desired gauge group, having the correct assignment of $U(1)$ charges.  This new two-form and the corresponding divisor should be understood in terms of a certain unified group, and from which the conventional $SU(5)$ gauge coupling unification relation is achieved if no flux is turned on the $U(1)$ part. 

We hope that this analysis provides a complete proof to the SM singularity suggested before: Either by relating to the spectral cover in the heterotic dual limit and explicit calculation of the charges of the matter fields. Gauge coupling unification can be another good clue for the model building, which is not shared by other models in the similar context having an intermediate Grand Unification. We may apply this method to a direct construction of the Standard Model in the native F-theory context. Mathematically, the appearance of an extra $U(1)$ as an element of Cartan subalgebra  in a larger unified group is very suggestive, so it would be interesting to extend the work to find more systematic method to find groups involving multiple $U(1)$'s.

\subsection*{Acknowledgements}
The author is grateful to Ralph Blumenhagen, Thomas Grimm, Stefan Groot-Nibbelink, Hirotaka Hayashi, Seung-Joo Lee, Christopher Mayrhofer, and Timo Weigand for discussions and correspondences. This work is partly supported by the National Research Foundation of Korea with grant number 2012-R1A1A1040695.

\end{document}